\documentclass[sn-mathphys]{sn-jnl}

\jyear{2023}%

\begin{document}

\title[A Workflow Management System Guide]{A Workflow Management System Guide}


\author[1,2]{\fnm{Caspar} \sur{Schmitt}}\email{Caspar.Schmitt@physik.uni-muenchen.de}

\author*[1]{\fnm{Boyang} \sur{Yu}}\email{Boyang.Yu@physik.uni-muenchen.de}

\author[1]{\fnm{Thomas} \sur{Kuhr}}\email{Thomas.Kuhr@lmu.de}

\affil*[1]{\orgdiv{Fakultät für Physik}, \orgname{Ludwig-Maximilians-Universität München}, \orgaddress{\street{Am Coulombwall 1}, \city{Garching}, \postcode{85748}, \country{Germany}}}

\affil[2]{\orgdiv{Werner-Heisenberg-Institut}, \orgname{Max-Planck-Institut für Physik}, \orgaddress{\street{Föhringer Ring 6}, \city{München}, \postcode{80805}, \country{Germany}}}


\abstract{
A workflow describes the entirety of processing steps in an analysis, such as employed in many fields of physics. Workflow management makes the dependencies between individual steps of a workflow and their computational requirements explicit, such that entire workflows can be executed in a stand-alone manner. Though the use of workflow management is widely recommended in the interest of transparency, reproducibility and data preservation, choosing among the large variety of available workflow management tools can be overwhelming. We compare selected workflow management tools concerning all relevant criteria and make recommendations for different use cases.
}




\maketitle

\section{Introduction}
\label{introduction}

The extraction of scientific results from data often requires many steps of data processing.
Usually there is a script for each step and the analyzer has to make sure they are executed in the right order.
This is error-prone and makes it difficult to reproduce results and preserve analyses.
Moreover it requires considerable manual work, in particular if the entire processing chain or parts of it have to be rerun many times which is often the case for analyses in particle physics, hadron and nuclear physics, or astrophysics.
These problems are addressed by making dependencies between processing steps explicit with workflow management systems.
They can automate the execution of processing steps and thus reduce the effort analyzers have to spend on the management of their workflows. Furthermore, workflow management make analyses stand-alone executables in the interest of transparency, reproducibility  and data preservation.

As workflows are relevant in many areas, various technical solutions exist.
For example, ref.~\cite{pipelines} lists several of them.
Given the large number of technical solutions it is hard to decide which of them should be chosen for which use case.
This article discusses criteria for the assessment of workflow management systems in Section~\ref{criteria}.
An example workflow is introduced in Section~\ref{example}.
The example workflow is implemented and the criteria are evaluated for a selected set of solutions in Section~\ref{assessment}.
The choice of system has to be made case by case as the relevance of criteria is use case dependent.
Some recommendations for typical use cases are given in Section~\ref{recommendations}.

We here assess five \textbf{workflow management frameworks} (Luigi, Snakemake, Yadage, Nextflow and CWL) which construct workflows, and a \textbf{workflow management platform} (Reana) that allows for additional tools to instantiate and run the workflows on remote clouds. All the tools discussed here are general-purpose which have the potential to be applied in various research fields and computing environments.

\section{Evaluation Criteria}
\label{criteria}
\textsc{User Interface}
\begin{itemize}
\item \textbf{Workflow language}:
Is the definition of a workflow given in a custom or standard language?
\item \textbf{Relation to analysis code}:
Does the description of the workflow factorize from the analysis code or is it integrated in the analysis code?
A tight integration can be beneficial if workflows are well defined and strongly coupled to the analysis code.
A factorized approach provides more flexibility and is particularly useful to implement existing code, but may require synchronized changes of workflow definition and analysis code.
Does the system support multiple languages?
\item \textbf{Boilerplate code}:
How much code is needed on average to include a step and its dependencies in a workflow?
\item \textbf{Visualization and monitoring}:
How can the dependency graph be visualized and the progress of execution be monitored?
Can partial results be inspected?
\item \textbf{Learning curve}:
How much effort does a beginner have to invest to get a simple/complex workflow running?
\end{itemize}
\newpage
\noindent\textsc{Features}
\begin{itemize}
\item \textbf{Supported programming languages}:
Is the system agnostic to the programming language used for the implementation of steps or does it only support specific ones?
\item \textbf{Data formats}:
Does the system support only specific input or output data formats?
Does it have features for the inspection of specific data formats?
\item \textbf{Dependency management}:
Does the system support one-to-many, many-to-one, and many-to-many dependencies?
Can dependencies be generated dynamically?
Are conditional dependencies supported?
Can loops be implemented?
\item \textbf{Execution control}:
Does it allow to pause and resume the execution?
Can changes be applied during the execution?
Can workflows be run top-down or up to specified target files? Can intermediate results be reused?
\item \textbf{Error handling}:
Does the system reliably detect error conditions in processing steps and what error handling/recovery strategies are supported?
\item \textbf{Logging and provenance}:
What means are provided to identify reasons for failures? Does the system keep track of when, where, and in which environment steps were executed and which output they produced?
\item \textbf{Version control and archivability}:
Are tools for version control and archivability implemented? Is external version control possible (e.g. using git)?
\item \textbf{Scalability}:
How easy or hard is it to change from a small test setup to a large production?
Can steps be executed in parallel?
\end{itemize}
\textsc{Resource Integrations}
\begin{itemize}
\item \textbf{Software and environment management}:
How are dependencies of the analysis code on operating systems and external libraries handled?
\item \textbf{Storage systems support}:
What kind of storage systems/protocols for input and output data are supported?
\item \textbf{Remote execution system support}: 
What kind of batch systems for the execution of steps are supported?
To what level of detail can batch system slot requirements be specified?
\item \textbf{Authentication and authorization mechanisms}:
What mechanisms of authentication and authorization for the access to resources are supported?
\end{itemize}
\textsc{Installation and Configuration}
\begin{itemize}
\item \textbf{Installation}:
How easy or hard is the installation of the system?
Does it require root access or certain tools?
\item \textbf{Architecture}:
Is it a single application or a client-server architecture?
Is the execution correctly resumed if the (client) application is terminated and restarted later?
\item \textbf{State management}:
How does the system keep track of the state of execution of a workflow?
In a \textit{report based} system the state of execution is stored in a central report file.
An advantage is that all information is at one place.
It come with the risk that the report file may not reflect the true state.
In a \textit{target based} system the state is determined dynamically by checking for the output of processing steps.
This avoids the problem of inconsistent information, but leads to a higher load on the systems that are queried for the state.
\item \textbf{Portability}:
Can a workflow easily be executed from different locations?
\end{itemize}
\textsc{Support and Management of the Tool}
\begin{itemize}
\item \textbf{Documentation}: What kind of documentation is provided and how useful is it?
\item \textbf{Support}:
Which communication channels or tools for support requests are provided?
What is the response time and quality to support requests?
Is there a large community support (e.g. on stackoverflow.com)?
How likely is it to get support from a specific science community?
\item \textbf{Tool developers}:
Who are the developers of the system?
Is it a single person, a small team, a large community, an institute, or a company?
\item \textbf{History and project activity}:
Since when does the project exist?
How actively is it developed further?
What is the frequency of commits or releases?
\item \textbf{User community}:
How large/diverse is the user community?
Is the system an established solution in some communities?
\item \textbf{Long term perspective}:
How likely is it that the system is still actively supported in several years?
\item \textbf{Lock-in}:
How hard or easy is it to change to a different workflow management system?
\item \textbf{License}:
What is the license of the product?
\item \textbf{Use in PUNCH}:
In which research fields is the workflow management system used within particle, astro-, astroparticle, hadron and nuclear physics (PUNCH)?
\end{itemize}

\section{Example Workflow}
\label{example}
For each system, we implement a simple example workflow, covering the key features. It consists of three processing steps:
\begin{itemize}
    \item \textbf{Task 1}: Generate a random seed and write it to a file.
    \item \textbf{Task 2}: Use the random seed to generate a given number, $N$, of random numbers and write them to a file.
    This step is executed two times, for $N=10$ and $N=5$.
    \item \textbf{Task 3}: Concatenate the files with random numbers, generated by calls of step 2.
\end{itemize}
This example workflow illustrates many features of workflow management:
\begin{itemize}
    \item Constructing processing steps and specifying dependencies between them to connect them in a workflow.
    \item One-to-many and many-to-one dependencies, as well as steps without dependencies.
    \item Passing parameters to and between processing steps.
    \item Using shell commands as well as python scripts in individual processing steps.
    \item Defining the input and output directory and file structure.
    \item Demonstrating the available workflow visualization tools.
    \item Specifying computational requirements, constructing virtual environments and running individual processing steps in them.
    \item Executing the entire workflow in a stand-alone manner.
\end{itemize}
For simplicity, the same random seed is utilized for each call of step 2. Consequently, the generated random number sequence remains consistent throughout. Although this may not be ideal for real-world scenarios, it does not impact the comparisons presented in this document.

\section{Assessment of Selected Workflow Management Systems}
\label{assessment}
\subsection{Workflow management frameworks}
Workflow management frameworks are used to construct workflows which automate task scheduling based on dependencies between tasks. We here discuss the Luigi, Snakemake, Yadage, Nextflow and CWL frameworks. The selection of reviewed workflow management frameworks is influenced by the authors' background and can be extended in future.

\subsubsection*{Luigi\footnote{https://github.com/spotify/luigi/}}
Luigi is a workflow management framework initially designed for uses in industry, with extensive support for dynamic workflow visualization and remote execution. The dependency logic is decentralized in Python classes with integrated analysis code.\\

\begin{figure}[h]
    \centering
    \begin{minipage}{0.65\textwidth}
        \includegraphics[width=\textwidth]{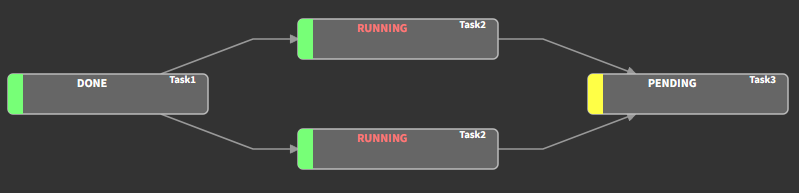}
    \end{minipage}\hfill
    \begin{minipage}{0.34\textwidth}
        \includegraphics[width=\textwidth]{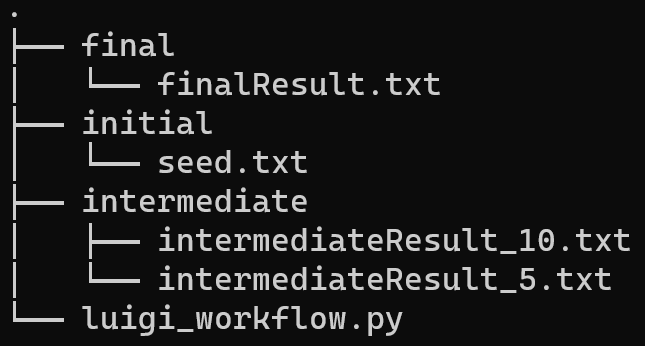}
    \end{minipage}
    \caption{Directed Acyclic Graph (DAG) forwarded by Luigi scheduler and directory structure for simple workflow in Luigi.}
    \label{fig:demoLuigi}
\end{figure}

Figure \ref{fig:demoLuigi} shows the example workflow implemented in a Luigi workflow. Task 1 generates the random seed and stores it in the file \texttt{initial/initialResult.txt}. It is used by Task 2 to generate an amount of random numbers, specified by a Luigi parameter, inside a docker container spawned from a given image. Task 3 concatenates the outputs of the calls of Task 2. Dependencies are defined in the \texttt{requires} method of task classes. For shell commands, the subprocess library has to be called.

\begin{python}
#luigi_workflow.py
import luigi
import subprocess, os
from luigi.contrib.docker_runner import DockerTask

class Task3(luigi.Task):
    def requires(self):
        return [Task2(NumberOfRandoms = 10), Task2(NumberOfRandoms = 5)]
    def output(self):
        return luigi.LocalTarget("final/finalResult.txt")
    def run(self):
        command = ["cat"]
        for input in self.input(): command.append(input.path)
        with self.output().open("w") as output:
            output.write(subprocess.check_output(command).decode())

class Task2(DockerTask):
    NumberOfRandoms = luigi.IntParameter()
    def requires(self):
        return Task1()
    def output(self):
        return luigi.LocalTarget(f"intermediateResult_{self.NumberOfRandoms}.txt")
    @property
    def binds(self):
        return [f"{os.getcwd()}:/workdir"]
    @property
    def docker_url(self):
        return 'unix://var/run/docker.sock'
    @property
    def image(self):
        return "container:v1"
    @property
    def command(self):
        return f"python3 random_numbers.py {self.requires().output().path} {self.output().path} {self.NumberOfRandoms}"

class Task1(luigi.Task):
    def output(self):
        return luigi.LocalTarget("initial/initialResult.txt")
    def run(self):
        with self.output().open("w") as output:
            output.write("42")

if __name__ == "__main__":
    luigi.build([Task3()], workers=2, local_scheduler=True)
\end{python}
\begin{python}
#random_numbers.py
import sys
import numpy

def generate_random_numbers(seed_file,output_file,NumberOfRandoms):
    with open(seed_file,"r") as seed:
        numpy.random.seed(int(seed.readlines()[0]))

    with open(output_file,"w") as output:
        for i in range(int(NumberOfRandoms)): 
            output.write(f"{numpy.random.random()}\n")

generate_random_numbers(*sys.argv[1:])
\end{python}
We call the workflow from the command line using \texttt{luigi --module Task3 --workers 2 --local-scheduler} or from Python code using \texttt{python3 luigi\_workflow.py}. The docker image is built from a Dockerfile (\texttt{docker build -t container:v1}) and bound to the working directory.
\begin{python}
#Dockerfile
FROM python:3.8-slim-buster

RUN pip install --no-cache-dir --upgrade pip \
  && pip install --no-cache-dir numpy

RUN mkdir /workdir
WORKDIR /workdir
\end{python}

\noindent\textsc{User Interface}
\begin{itemize}
\item \textbf{Workflow language}: Python.
\item \textbf{Relation to analysis code}: For each task, the analysis code is integrated into the workflow logic.
\item \textbf{Boilerplate code}: Minimal. Variable values can be passed directly between tasks by employing Luigi parameters, with automatic handling of conversion of Python types. Typically per task class: one function each for input, output and analysis code.
\item \textbf{Visualization and monitoring}: Workflow monitoring via local scheduler or in \textit{Central Scheduler} which runs in daemon mode. The latter allows for excellent dynamic DAG visualization and optional dynamic status messages. Partial results can be inspected manually for running and completed tasks.
\item \textbf{Learning curve}: Straight-forward setup of workflows that factorize into targets and tasks, which are expressed in Python classes with integrated analysis code. Analysis scripts need to be altered to accomodate workflow logic code.
\end{itemize}
\textsc{Features}
\begin{itemize}
\item \textbf{Supported programming languages}: Python.
\item \textbf{Data formats}: Any format. Tasks are marked as complete once output files are created and written successfully, without further inspection. Output directories are created automatically.
\item \textbf{Dependency management}: Decentralized and integrated in analysis code. Task dependencies are explicit by requiring previous tasks. For dependency inspection, workflows can be run dryly without executing tasks. Support for one-to-many, many-to-one, and many-to-many dependencies. If the directed acyclic graph (DAG) cannot be fully built a priori, such as for dynamic and conditional dependencies, tasks can be yielded from within another task. Loops can be implemented by using task parameters.
\item \textbf{Execution control}: Execution cannot be paused, but workflows can be run up to specified intermediate tasks, as well as top-down. Intermediate results are reused. Changes require a restart of the workflow. Workflows can be run from command line or Python code. Multiple tasks can be batched together and scheduled as one. Built-in functionality to solve the atomic write problem (e.g. creation of temporary directories and move commands). For easy customizability, extensive configuration files can be added. For each task, further resources can be specified. Tasks can be prioritised. Upon successful completion of a workflow, dedicated code can be executed (e.g. sending a completion message).
\item \textbf{Error handling}: The workflow is marked as failed if individual tasks fail, unless multiple retries are required. Failed tasks are reported, along with their parameters and error messages produced during their execution, but (incomplete) output is not cleaned up. Built-in event system allows to register callbacks to events and trigger them for a specific task class (e.g. on success or failure of a task).
\item \textbf{Logging and provenance}:
\textit{Central Scheduler} captures execution history of tasks. Luigi allows for the specification of log files with adjustable log level, limited provenance information.
\item \textbf{Version control and archivability}: External version control possible.
\item \textbf{Scalability}: Simple scalability. Number of so-called workers can be specified. For multiple workers, tasks will be automatically run in parallel, whenever possible. No execution is transferred, i.e. workers schedule and execute all tasks. This may limit scalability eventually.
\end{itemize}
\textsc{Resource Integrations}
\begin{itemize}
\item \textbf{Software and environment management}: Docker containers can be run as tasks. Luigi analysis workflow\footnote{https://github.com/riga/law} (law), which is built on top of Luigi, provides extensive features for environment sandboxing on task level. Support within commercial cloud systems.
\item \textbf{Storage systems support}: Target classes map to files on remote locations (e.g. via ssh or ftp, on the Apache Hadoop Distibuted File System etc.).
\item \textbf{Remote execution system support}: LSF batch system support. HTCondor and LHC Computing Grid support can be implemented (see e.g. b2luigi\footnote{https://github.com/nils-braun/b2luigi} or law). Support for Jobs on Sun Grid Engine, Spark Jobs etc. Cloud execution support Kubernetes, Dropbox, Google Cloud, Salesforce, Amazon AWS Cloud etc. Extensive support for Apache Hadoop jobs and the Apache Hadoop Distributed File System (HDFS). Tasks can be marked as local, suppressing a batch submission. No support for distribution of execution.
\item \textbf{Authentication and authorization mechanisms}: Luigi allows to access environment variables, which may contain authentification tokens.
\end{itemize}
\textsc{Installation and Configuration}
\begin{itemize}
\item \textbf{Installation}: Python2 or Python3 is necessary for using Luigi. Luigi can be easily installed from the Python Package Index (pip), without root access rights.
\item \textbf{Architecture}: Single application. Upon restart, existing output files will be detected and corresponding tasks will not be re-run. Running batch jobs are automatically detected and not re-submitted.
\item \textbf{State management}: Target based.
\item \textbf{Portability}: Workflows can be executed from anywhere by pointing to the corresponding scripts, as long as file paths are correctly specified.
\end{itemize}
\textsc{Support and Management of the Tool}
\begin{itemize}
\item \textbf{Documentation}: Extensive documentation on official website. Open source code on Github.
\item \textbf{Support}: Support on stackoverflow.com. Luigi was initially intended for Spotify but translates to many other industries.
\item \textbf{Tool developers}: Spotify Group.
\item \textbf{History and project activity}: Developed by and for the Spotify group, open sourced in 2012 and in active development  (currently multiple new commits monthly).
\item \textbf{User community}: Active community, frequently used by companies.
\item \textbf{Long term perspective}: Good, given prevalence, active development and industrial use.
\item \textbf{Lock-in}: Likely, since analysis code structure needs to be modified significantly to accomodate workflow logic and no support for other workflow management frameworks.
\item \textbf{License}: Luigi is licensed under the Apache 2.0 License with free permission to alter it in any way.
\item \textbf{Use in PUNCH}: CMS (Law), Belle II Experiment (b2luigi).
\end{itemize}

\subsubsection*{Snakemake\footnote{https://github.com/snakemake/snakemake}}
Snakemake is a workflow management framework designed for uses in research, with extensive support for environment management and remote execution. It features a very simple Python-based syntax and accomodates shell commands and external scripts in multiple programming languages with minimal adaptations. Due to this and its support for other worklow management frameworks, lock-in is very unlikely. All workflow logic is centralized in the so-called \textit{snakefile}, similar to a Makefile.\\

\begin{figure}[h]
    \centering
    \begin{minipage}{0.45\textwidth}
        \includegraphics[width=\textwidth]{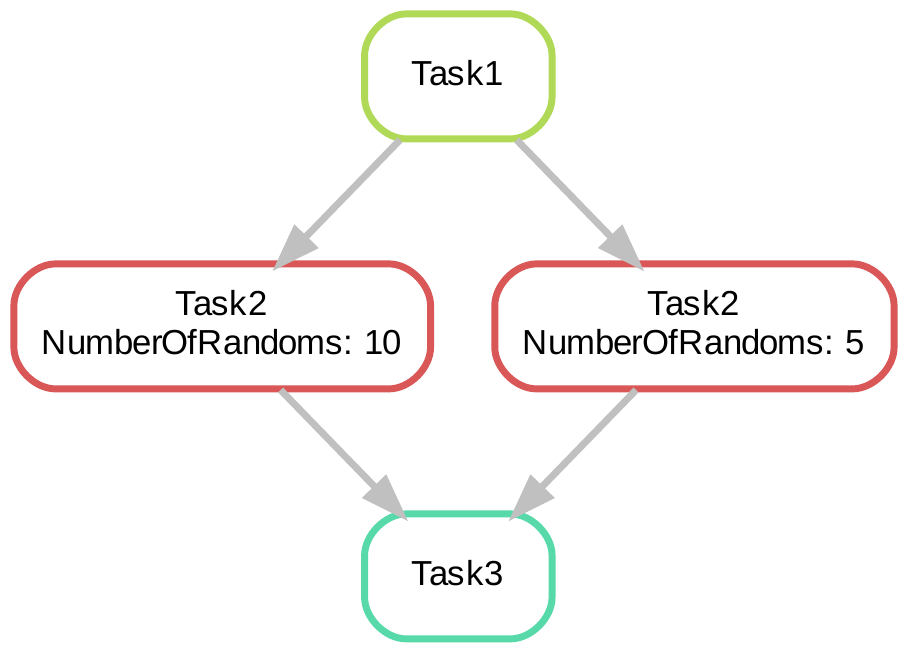}
    \end{minipage}\hfill
    \begin{minipage}{0.45\textwidth}
        \includegraphics[width=\textwidth]{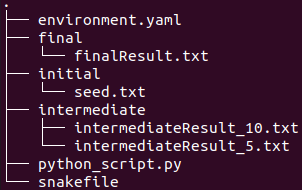}
    \end{minipage}
    \caption{Directed Acyclic Graph (DAG) generated with $\texttt{--dag \string| dot \string| display}$ flag and directory structure for simple workflow in Snakemake.}
    \label{fig:demoSnake}
\end{figure}

Figure \ref{fig:demoSnake} illustrates the implementation of the example workflow. Task 1 generates the random seed. It is used by Task 2 to generate random numbers, by calling a separate python script. Snakemake creates an environment for the script to run in, which includes the specified packages on any machine. This can be achieved via the integrated conda package manager and within a container spawnable from a given docker image. With each call of Task 2, we pass a wildcard parameter to the script, which specifies how many numbers are generated. Task 3 concatenates the outputs of the calls of Task 2.
Dependencies between tasks are implicitly defined by their output and input files, like in a Makefile.

\begin{python}
#snakefile
rule Task3:
    input:
        "intermediate/intermediateResult_10.txt",
        "intermediate/intermediateResult_5.txt"
    output:
        "final/finalResult.txt"
    shell:
        "cat {input} > {output}"

rule Task2:
    input:
        "initial/seed.txt"
    output:
        "intermediate/intermediateResult_{NumberOfRandoms}.txt"
    conda:
        "environment.yaml"
    containerized:
        "container.sif"
    params:
        current_NumberOfRandoms = lambda wildcards: int(wildcards.NumberOfRandoms)
    script:
        "python_script.py"

rule Task1:
    output:
        "initial/seed.txt"
    shell:
        "echo 42 > {output}"
\end{python}

\begin{python}
#python_script.py
input_file = snakemake.input[0]
output_file = snakemake.output[0]
number_of_randoms = snakemake.params.current_NumberOfRandoms

import numpy
with open(input_file,"r") as input:
    numpy.random.seed(int(input.readlines()[0]))
with open(output_file,"w") as output:
    for i in range(number_of_randoms): output.write(f"{numpy.random.random()}\n")
\end{python}
\begin{python}
#environment.yaml
name: environment
dependencies:
    - numpy
\end{python}

We call the workflow from the command line with $\texttt{snakemake --cores 2 --use-singularity --use-conda}$, where \texttt{cores} specifies the number of parallel CPU cores. Alteratively for supported cluster execution, use $\texttt{snakemake --cluster qsub --jobs 2}$.\\

\noindent\textsc{User Interface}
\begin{itemize}
\item \textbf{Workflow language}: Custom human readable Python-based language with additional syntax to define tasks and workflow specific properties.
\item \textbf{Relation to analysis code}:  The description of the workflow (\textit{snakefile}) factorizes entirely from the analysis code (separate scripts).
\item \textbf{Boilerplate code}: Minimal. Typically per task in \textit{snakefile}: one line each for input, output, script or shell command and additional arguments. Minimal changes to separate scripts allow to pass arguments directly, which makes parsing code unnecessary.
\item \textbf{Visualization and monitoring}: Built-in commands to print workflow visualization to file (option to print entirety of workflow or omit multiple calls of tasks). Dynamic workflow monitoring through panoptes\footnote{https://github.com/panoptes-organization/panoptes} server and terminal-printout, albeit no dynamic DAG. Partial results and log files can be inspected manually for running and completed tasks.
\item \textbf{Learning curve}: Quick and straight-forward setup of workflows. So-called \textit{snakefile} contains all workflow logic with all workflow tasks (called \textit{rules}).
\end{itemize}
\textsc{User Interface}
\begin{itemize}
\item \textbf{Supported programming languages}: Separate scripts and shell commands can be called directly. Support for Shell, Python, R \& R Markdown, Julia, Rust and Jupyter notebooks for interactive development.
\item \textbf{Data formats}: Any format. Tasks are marked as complete once output files are created and written successfully. Output file inspection possible through non-emptyness and checksum compliance. Output directories are created automatically. Specified temporary output files are deleted once not needed anymore. Specified protected output files are write-protected. Alternatively, directories can be specified as outputs.
\item \textbf{Dependency management}: Centralized. \textit{Snakefile} contains inputs and outputs for all tasks, from which their dependencies are determined automatically and motivated in printout. Instead of via filenames, dependencies can also be required directly by referring to outputs of specific tasks. Ambiguities in dependencies (e.g. tasks with identical outputs) can be resolved by explicitely specifying the executing order for tasks. For dependency inspection, workflows can be run dryly without executing tasks. Support for one-to-many, many-to-one, and many-to-many dependencies. Dynamic and conditional dependencies can be implemented by marking tasks as checkpoints, for which the directed acyclic graph (DAG) is re-evaluated at run-time based on the output of previous tasks. Loops can be implemented by using wildcard parameters.
\item \textbf{Execution control}: Execution cannot be paused, but workflows can be run up to specified intermediate outputs, as well as top-down. Intermediate results are reused, unless the user requires a re-run. Changes require a restart of the workflow. Execution flags can be stored in profiles. For easy customizability, parameters and initial input files can be specified in configuration file. Built-in functionality to benchmark tasks. For each task, the number of threads and further resources can be specified. Tasks can be prioritised. Upon successful completion of a workflow, dedicated code can be executed (e.g. sending a completion message).
\item \textbf{Error handling}: The workflow is marked as failed if individual tasks fail, unless multiple retries are required. Failed tasks are reported, along with their parameters and error messages produced during their execution. (Incomplete) outputs of failed tasks are deleted automatically.
\item \textbf{Logging and provenance}:
Snakemake allows for the specification of log files for each task. Additionally global log files for each workflow execution are written, including the executed tasks with their parameters, timestamps and activated environments.
\item \textbf{Version control and archivability}: Snakemake tracks the code that was used to create output files; Tasks can be re-run automatically for changes in corresponding code. External version control possible.
\item \textbf{Scalability}: Simple scalability. Modularization in sub-workflows possible. Number of CPU cores can be specified. For multiple cores, tasks will be automatically run in parallel, whenever possible. Large workflows can be executed in batches. Support for scatter-gather workflows.
\end{itemize}
\textsc{Resource Integrations}
\begin{itemize}
\item \textbf{Software and environment management}: For fully reproducible workflows, software tools and libraries can be speficied in isolated software environments (globally or for individual tasks). This can be achieved using the integrated conda package management or by running within a (docker) container spawned from a given image. Workflows can then be executed without additional prerequisites, while software packages are supplied automatically on any machine. Snakemake wrapper repository provides frequently used scripts.
\item \textbf{Storage systems support}: Snakemake allows to retrieve and upload files from and to several remote locations (e.g. via ssh, http and commercial cloud services).
\item \textbf{Remote execution system support}: LSF batch system support. HTCondor and LHC Computing Grid support can be implemented (e.g. analogous to b2luigi). Cluster execution support for cluster engines that support shell scripts and have access to a common filesystem, with extensive job properties. Support for Distributed Resource Management Application API (DRMAA) and Slurm. Tasks can be assigned to groups, which are submitted together to the same computing node. Tasks can be marked as local, suppressing a batch submission. Cloud execution support (Kubernetes via Google cloud engine, Google Cloud Life Sciences with GPUs, Tibenna on Amazon Web Services, GA4GH TES).
\item \textbf{Authentication and authorization mechanisms}: Snakemake allows to access environment variables, which may contain authentification tokens.
\end{itemize}
\textsc{Installation and Configuration}
\begin{itemize}
\item \textbf{Installation}: Python3 is necessary for using Snakemake. A full version of Snakemake can be easily installed using Conda and Mamba, without root access rights. Similarly, a minimal version depending only on bare necessities is available.
\item \textbf{Architecture}: Single application. Upon restart, existing output files will be detected and corresponding tasks will no be re-run $\Rightarrow$ \textit{Exception}: If a task features input files with newer timestamps than the existing output files, it will be re-run.
\item \textbf{State management}: Target based.
\item \textbf{Portability}: Workflows can be executed from anywhere by pointing to the \textit{snakefile}, as long as file paths are correctly specified. Tasks marked as \textit{shadow rules} are run in isolated temporary directories. Workflows and their outputs can be chained.
\end{itemize}
\textsc{Support and Management of the Tool}
\begin{itemize}
\item \textbf{Documentation}: Excellent and extensive documentation on official website. Open source code on Github.
\item \textbf{Support}: Extensive support on stackoverflow.com and discord. Snakemake was initially intended for uses in bio-informatics but translates to any research field.
\item \textbf{System developers}: Small team of developers led by Dr. Johannes Köster (University of Duisburg-Essen, Germany).
\item \textbf{History and project activity}: Snakemake exists since 2012 with ongoing active development (currently multiple new commits monthly).
\item \textbf{User community}: Very active community with $>7$ new citations per week.
\item \textbf{Long term perspective}: Good, given prevalence and active development.
\item \textbf{Lock-in}: No. Snakemake features built-in tools to export workflows to \textit{Common Workflow Language}\footnote{https://www.commonwl.org/} (CWL), which serves as comprehensive standard for workflow management frameworks. Furthermore, the execution of individual tasks can be handed over to other workflow management frameworks. Separate scripts are only minimally changed for integration into Snakemake.
\item \textbf{License}: Snakemake is licensed under the MIT License with free permission to alter it in any way.
\item \textbf{Use in PUNCH}: LHCb Experiment, Radioastronomy.
\end{itemize}

\subsubsection*{Yadage\footnote{https://github.com/yadage/yadage}}
Yadage is a workflow management software designed for scientific research that offers support for containerization, remote execution, and environment management. It features a YAML-based syntax that is easy to edit and allows users to include shell commands, external scripts, and various programming languages in their workflows. Yadage provides a flexible and scalable solution for researchers to manage parameterised and reproducible workflows, and is also designed to integrate with other workflow engines by centralizing all workflow logic in the YAML-based workflow definition file.\\

\begin{figure}[H]
    \centering
    \begin{minipage}{0.34\textwidth}
        \includegraphics[width=\textwidth]{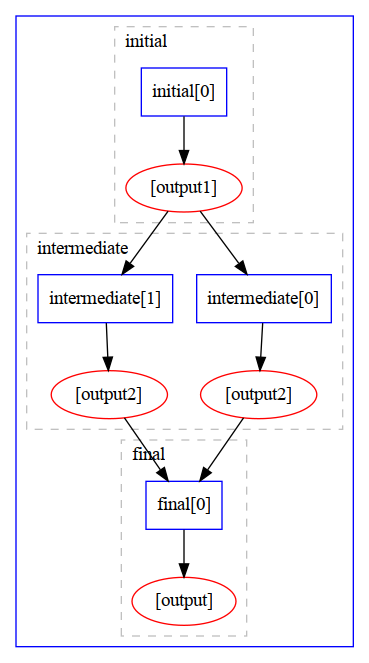}
    \end{minipage}\hfill
    \begin{minipage}{0.51\textwidth}
        \includegraphics[width=\textwidth]{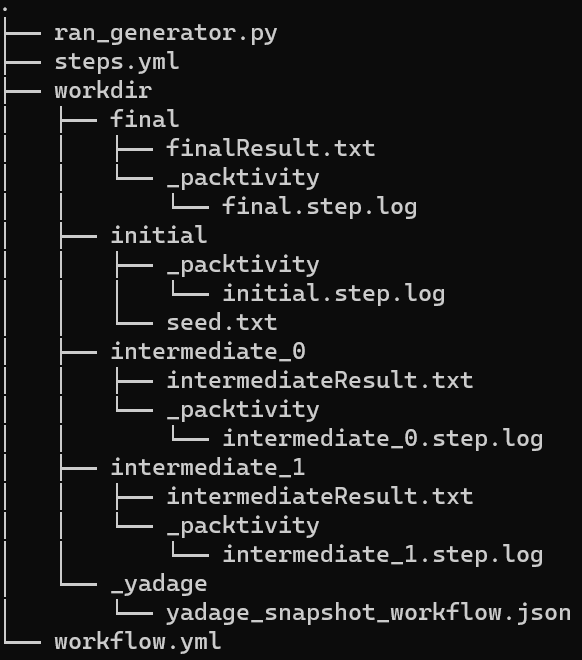}
    \end{minipage}
    \caption{Directed Acyclic Graph (DAG) generated with $\texttt{--visualize}$ flag and directory structure for simple workflow in Yadage.}
    \label{fig:demoYadage}
\end{figure}

Figure \ref{fig:demoYadage} shows the implementation of the example workflow. Task 1, initial stage, generates a random seed. It is used by Task 2, intermediate stage, to generate random numbers, by calling a separate python script. Task 3 concatenates the outputs of the calls of Task 2. The random seed and numbers of random numbers can be either hard-coded as shown in the example or given as inputs. The whole workflow is stored in the main file \textit{workflow.yml} while the concrete operation of each stage is defined in the packtivity file \textit{steps.yml}. The dependencies have to be explicitly specified both between the stages, in \textquotedblleft dependencies\textquotedblright \ part, and between the outputs of stages, in \textquotedblleft scheduler - parameters - inputfile\textquotedblright \ part. The execution of Task 2 in a container was however not successful for both local and online container images. This issue has been reported to the developers.

\begin{python}
#yadage-main: workflow.yml
stages:
- name: initial
  dependencies: []
  scheduler:
    scheduler_type: singlestep-stage
    parameters:
      seed: 42
      outputfile: '{workdir}/seed.txt'
    step: {$ref: 'steps.yml#/seedwriter'}

- name: intermediate
  dependencies: [initial]
  scheduler:
    scheduler_type: multistep-stage
    parameters:
      inputfile: {step: initial, output: output1}
      numberOfRandoms: [10, 5]
      outputfile: "{workdir}/intermediateResult.txt"
    scatter:
      method: zip
      parameters: [numberOfRandoms]
    step: {$ref: 'steps.yml#/randomgenerator'}

- name: final
  dependencies: [intermediate]
  scheduler:
    scheduler_type: singlestep-stage
    parameters:
      inputfiles: {steps: intermediate, output: output2}
      outputfile: "{workdir}/finalResult.txt"
    step: {$ref: 'steps.yml#/resultscombiner'}
\end{python}

\begin{python}
#yadage-packtivity: steps.yml
seedwriter:
  process:
    process_type: 'string-interpolated-cmd'
    cmd: echo {seed} > {outputfile}
  environment:
    environment_type: 'localproc-env'
  publisher:
    publisher_type: 'interpolated-pub'
    publish:
      output1: '{outputfile}'

randomgenerator:
  process:
    process_type: 'string-interpolated-cmd'
    cmd: ./../../ran_generator.py {inputfile} {numberOfRandoms} {outputfile}
  environment:
    environment_type: 'localproc-env'
  publisher:
    publisher_type: 'interpolated-pub'
    publish:
      output2: '{outputfile}'

resultscombiner:
  process:
    process_type: 'string-interpolated-cmd'
    cmd: cat {inputfiles} > {outputfile}
  environment:
    environment_type: 'localproc-env'
  publisher:
    publisher_type: 'interpolated-pub'
    publish:
      output: '{outputfile}'
\end{python}

\begin{python}
#random generator for intermediate stage: ran_generator.py
#!/usr/bin/env python
import numpy
import sys

input_file = str(sys.argv[1])
number_of_randoms = int(sys.argv[2])
output_file = str(sys.argv[3])

with open(input_file, "r") as input:
    numpy.random.seed(int(input.readlines()[0]))
with open(output_file, "w") as output:
    for i in range(number_of_randoms):
        output.write(str(numpy.random.random())+'\n')
\end{python}

We call the workflow from the command line with $\texttt{yadage-run workdir workflow.yml}$.\\

\noindent\textsc{User Interface}
\begin{itemize}
\item \textbf{Workflow language}: YAML conforming to pre-established workflow JSON schemas.
\item \textbf{Relation to analysis code}: The description of the workflow factorizes entirely from the analysis code (separate scripts). But it is also possible to integrate the analysis codes into the workflow script.
\item \textbf{Boilerplate code}: A lot. To include a new step, both the workflow script (to arrange the steps) and the packtivity script (to define the steps) must be updated with the necessary code.
\item \textbf{Visualization and monitoring}: Visualizations including one DAG and one GIF animation showing the evolution of the DAG over time are available after execution. No dynamic workflow monitoring during the execution. Partial results can be inspected manually by checking the log files or running the stages separately.
\item \textbf{Learning curve}: Hard to start or customise due to the lack of documentation and user community. 
\end{itemize}
\textsc{User Interface}
\begin{itemize}
\item \textbf{Supported programming languages}: The system is agnostic to the programming language used for the implementation of steps.
\item \textbf{Data formats}: Any formats. No specific features for the inspection of data formats.
\item \textbf{Dependency management}: Centralized in \textit{workflow.yml} where the dependencies are manually defined. One-to-many, many-to-one, and many-to-many dependencies are supported. Dynamical, conditional and looping dependencies can not be implemented.
\item \textbf{Execution control}: Yadage does not support pausing or resuming workflows, and it does not allow changes to be made during execution either. Workflows must be run in their entirety, from start to finish, and intermediate results can be reused by manually running certain stages. Yadage does not support running workflows in a top-down or target-driven fashion.
\item \textbf{Error handling}: The workflow will be marked as failed if any of its individual tasks fail, and error messages produced during the execution of failed tasks are reported. However, (incomplete) outputs of failed tasks are not deleted automatically.
\item \textbf{Logging and provenance}: The specification of log files is not allowed. Error messages with traceback information are recorded in the execution history logs.
\item \textbf{Version control}: The code used to create output files is tracked in log files, which also contain information about the execution history. External version control tools like git can also be used for additional tracking and management of code changes.
\item \textbf{Scalability}: Simple scalability. Parallel executions are supported on a Celery cluster or IPython clusters.
\end{itemize}
\textsc{Resource Integrations}
\begin{itemize}
\item \textbf{Software and environment management}: Docker containers can be run as a task.
\item \textbf{Storage systems support}: No known limitations for a single machine. A shared filesystem is needed for multi-machine distributed execution.
\item \textbf{Remote execution system support}: Celery clusters and IPython clusters are supported. Custom backends for other batch systems can be implemented by specifying the module holding the backend and proxy classes in the packtivity package.
\item \textbf{Authentication and authorization mechanisms}: Yadage allows to access environment variables, which may contain authentication tokens.
\end{itemize}
\textsc{Installation and Configuration}
\begin{itemize}
\item \textbf{Installation}: Yadage can be easily installed from the Python Package Index (pip), without root access rights. It's also possible to use the official Docker image without installation. 
\item \textbf{Architecture}: Single application. Upon restart, existing output repository will be detected and the task will not start unless allowing overwrite with an extra command line option. 
\item \textbf{State management}: Target based.
\item \textbf{Portability}: Workflows can be easily executed from different locations with the help of Dockers. 
\end{itemize}
\textsc{Support and Management of the Tool}
\begin{itemize}
\item \textbf{Documentation}: Documentation and tutorials are fragmentary, requiring users to rely heavily on examples and/or explore the source code on Github even for basic usage.
\item \textbf{Support}: Only Github is provided, without guaranteed response to bug reports. No community support. 
\item \textbf{System developers}: Small team of 3 developers led by Prof. Dr. Lukas Heinrich (Technical University Munich, Germany).
\item \textbf{History and project activity}: First release came in 2017, inactive development especially in the recent years.
\item \textbf{User community}: Small community with 14 citations in total.
\item \textbf{Long term perspective}: Bad, given the incomplete documentation, inactive development and tiny developing group.
\item \textbf{Lock-in}: No. The YAML-based framework can be easily transformed to a different workflow management system.
\item \textbf{License}: Yadage is licensed under the MIT License with free permission to alter it in any way.
\item \textbf{Use in PUNCH}: ALICE, ATLAS, CMS, LHCb.
\end{itemize}

\subsubsection*{Nextflow\footnote{https://github.com/nextflow-io/nextflow}}
Nextflow is a powerful workflow management framework designed for scientific research, with extensive support for containerization, environment management, and remote execution on various batch systems and cloud servers. Its Groovy-based DSL syntax is easy to use and accommodates shell commands, external scripts, and multiple programming languages. Nextflow enables reproducibility and scalability of workflows, while also providing secure management of sensitive information with its built-in secret module. It also offers a rich set of built-in operators that can largely simplify the code for complex workflows. The user community of Nextflow is strong, with a large number of pre-built workflows available for various research domains. While Nextflow is primarily intended for use in bio-informatics, it can be easily employed in workflows from other research fields as well.\\

\begin{figure}[H]
    \centering
    \begin{minipage}{0.28\textwidth}
        \includegraphics[width=\textwidth]{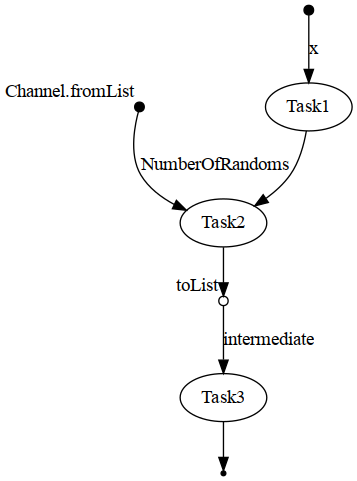}
    \end{minipage}\hfill
    \begin{minipage}{0.7\textwidth}
        \includegraphics[width=\textwidth]{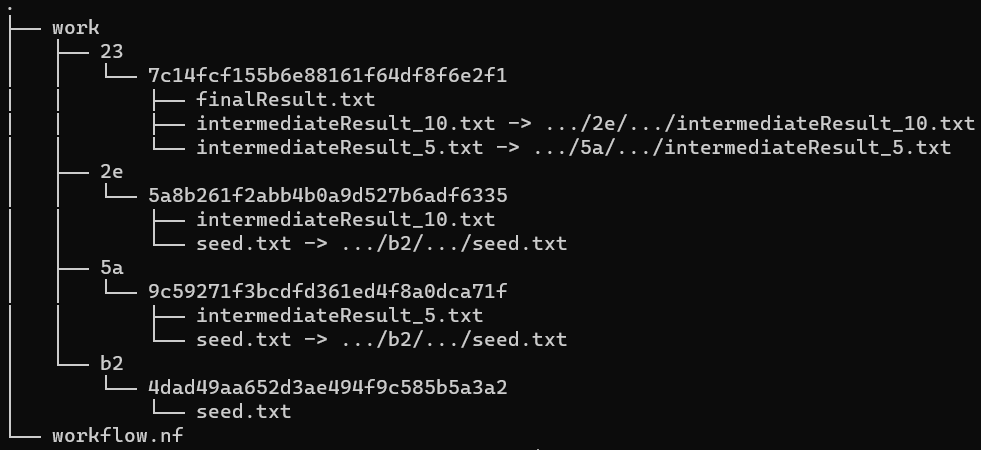}
    \end{minipage}
    \caption{Directed Acyclic Graph (DAG) generated with \texttt{-with-dag} flag and directory structure for simple workflow in Nextflow.}
    \label{fig:demoNextflow}
\end{figure}

Figure \ref{fig:demoNextflow} illustrates the realization of the example workflow. Task 1 generates a random seed using the input value \(x\) provided in the code. Task 2 utilizes this seed to generate random numbers by invoking an in-line Python script within the local container environment ``container:v1''. The number of random numbers to be generated is determined by loading values from a predefined list using the \(Channel.fromList\) function. Task 3 collects the outputs of Task 2's invocations through the \(toList\) function and concatenates the results using an in-line command line. The entire workflow is defined in the \textit{workflow.nf} file, and the dependencies are specified within the \(workflow\) module in the form of functions and arguments. Nextflow automatically creates separate directories for each sub-task during execution, eliminating the need for manual path specification.

\begin{python}
# workflow.nf
params.seed = 42
params.repeats = [10,5]

process Task1 {
  input:
    val x
  output:
    path 'seed.txt'
  "echo $x > seed.txt"
}

process Task2 {
  container 'container:v1'
  input:
    path 'seed.txt'
    each NumberOfRandoms
  output:
    path 'intermediateResult_*.txt'
  """
  #!/usr/bin/python
import numpy
with open('seed.txt', "r") as input:
  numpy.random.seed(int(input.readlines()[0]))
with open('intermediateResult_${NumberOfRandoms}.txt', "w") as output:
  for i in range($NumberOfRandoms):
    output.write(str(numpy.random.random()) + '\\n')
  """
}

process Task3 {
  input:
    path intermediate
  output:
    path 'finalResult.txt'
  "cat $intermediate > finalResult.txt"
}

workflow {
  NumberOfRandoms = Channel.fromList(params.repeats)
  initial = Task1(params.seed)
  intermediate = Task2(initial, NumberOfRandoms).toList()
  Task3(intermediate)
}
\end{python}

We call the workflow from the command line with $\texttt{nextflow run workflow.nf}$.\\

\noindent\textsc{User Interface}
\begin{itemize}
\item \textbf{Workflow language}:
Groovy-based Domain Specific Language.
\item \textbf{Relation to analysis code}:
The description of the workflow factorizes entirely from the analysis code (separate scripts). But it is also possible to integrate the analysis codes into the workflow script.
\item \textbf{Boilerplate code}:
Minimal. Similar to snakemake, only one line each for input, output, script or shell command and additional arguments are necessary. Managing channels and variables adds complexity, but on the other hand, can significantly simplify the script or shell command together with powerful built-in operators. 
\item \textbf{Visualization and monitoring}:
A wide range of built-in tracing and visualisation tools are provided, including execution log and report, trace report, timeline report and DAG visualisation. Dynamic workflow monitoring through Weblog via HTTP, albeit no dynamic DAG. Partial results and log files can
be inspected manually for running and completed tasks.
\item \textbf{Learning curve}:
Nextflow has a steeper learning curve compared to Luigi or Snakemake due to its complexity. However, the detailed tutorials and documentation make the learning process quicker, especially for users with a basic knowledge of programming.
\end{itemize}
\textsc{Features}
\begin{itemize}
\item \textbf{Supported programming languages}:
The system is agnostic to the programming language used for the implementation of steps.
\item \textbf{Data formats}:
Any formats. No specific features for the inspection of data formats.
\item \textbf{Dependency management}: Centralized in \textit{workflow.nf} where the dependencies are defined through channels and processes. One-to-many, many-to-one, and many-to-many dependencies are possible by leveraging process I/O and channel operators. The use of dynamic directives enables the dynamic generation of dependencies. Conditional dependencies and loops can be built through operators.
\item \textbf{Execution control}: The execution of all processes in the workflow is tracked by Nextflow. The workflow can be paused by e.g. killing the running process, and then resumed with the $\texttt{nextflow run workflow.nf -resume}$ command. Changes can also be made to the workflow before resuming. In the new execution, only the modified part of the workflow is re-executed, while the cache of other parts is utilized.
\item \textbf{Error handling}:
The workflow will be marked as failed if any of its individual tasks fail, unless multiple retries are required. Error messages produced during the execution of failed tasks are reported. However, (incomplete) outputs of failed tasks are not deleted automatically.
\item \textbf{Logging and provenance}:
The logging behavior of Nextflow can be customized by specifying fields of interest. In the event of errors, traceback information is captured in both execution logs and reports. Additionally, error handlers can be manually incorporated into the workflow to handle runtime or process errors that stop the workflow execution.
\item \textbf{Version control}:
The code used to create output files is tracked in log files, which also contain information about the execution history. Support for BitBucket, GitHub and GitLab are integrated with specialised functions and centralised credential management tools for additional tracking and management of code changes.
\item \textbf{Scalability}:
Simple scalability. Support parallel executions on multiple CPU cores or on a wide range of batch systems and cloud platforms.
\end{itemize}
\textsc{Resource Integrations}
\begin{itemize}
\item \textbf{Software and environment management}:
Nextflow has built-in support for Conda and Spack that enables the configuration of workflow dependencies using their recipes and environment files. This allows Nextflow applications to use popular tool collections such as Bioconda whilst taking advantage of the configuration flexibility provided by Nextflow. Containers built at different runtimes can be executed as a task.
\item \textbf{Storage systems support}:
Support both local and remote storage systems. The gap between cloud-native storage and data analysis workflows is bridge by Fusion which is a distributed virtual file system for cloud-native data workflows, optimised for Nextflow workloads.
\item \textbf{Remote execution system support}: 
Nextflow supports a large number of batch systems including Bridge, Flux, GA4GH TES, HyperQueue, HTCondor, Apache Ignite, LSF, Moab, NQSII, OAR, PBS/Torque, PBS Pro, SGE and SLURM, as well as cloud servers including AWS Batch, Azure Batch, Google Cloud, Google Life Sciences and Kubernetes.
\item \textbf{Authentication and authorization mechanisms}:
Nextflow provides a built-in secret module that manages authentication and authorization mechanisms in a centralized manner. This allows for decoupling the use of secrets in workflows from the workflow code and configuration files. By using this module, sensitive information such as passwords, access tokens, and encryption keys can be securely stored and managed separately from the workflow code. 
\end{itemize}
\textsc{Installation and Configuration}
\begin{itemize}
\item \textbf{Installation}:
Bash and Java are necessary for using Nextflow. Nextflow can be easily installed with \textit{wget}, \textit{curl} or from Conda without root access rights.
\item \textbf{Architecture}:
Single application. Upon restart, existing output files will be detected and corresponding tasks will not be re-run. But the jobs can be manually resumed with $\texttt{nextflow run workflow.nf -resume}$.
\item \textbf{State management}:
Target based.
\item \textbf{Portability}:
Workflows can be easily executed from different locations with the help of containers.
\end{itemize}
\textsc{Support and Management of the Tool}
\begin{itemize}
\item \textbf{Documentation}: 
Excellent and extensive documentation on official website. Open source code on Github.
\item \textbf{Support}:
Extensive support on Github discussions, stackoverflow.com and Slack chat. High quality Nextflow workflows from the community are collected and shared within the nf-core project. Yearly workshop showcasing researcher's workflows and advancements in the langauge are available on the Nextflow YouTube Channel. Nextflow is primarily designed and optimized for use in bio-informatics workflows. While it can be adapted for other research fields, support and resources may be limited. 
\item \textbf{Tool developers}:
A professional company, Seqera Labs, from Spain, funded by the Chan Zuckerberg Initiative and having deep cooperation with Amazon, Google and Microsoft.
\item \textbf{History and project activity}:
Nextflow exists since 2013 with highly active ongoing development (more than 10 commits weekly).
\item \textbf{User community}:
Very active and huge community, but limited to bio-informatics.
\item \textbf{Long term perspective}:
Very good perspective, given active development and community.
\item \textbf{Lock-in}:
No. There is no need to modify the individual analysis code before implementing Nextflow.
\item \textbf{License}:
Nextflow is licensed under the Apache 2.0 License with free permission to alter it in any way.
\item \textbf{Use in PUNCH}:
None.
\end{itemize}

\subsubsection*{Common Workflow Language\footnote{https://github.com/common-workflow-language/cwltool}}
CWL (Common Workflow Language) is an open standard for YAML-based workflow management, specifically designed for data-intensive scientific research. It provides researchers with a standardized and portable approach to describe computational tasks and their dependencies, supporting multiple programming languages. \\

Here we review $\texttt{cwltool}$, a powerful reference implementation that facilitates local execution of CWL workflows. Serving as a command line tool and Python library, $\texttt{cwltool}$ enables users to interpret CWL documents, manage inputs and outputs, and execute workflows on a local machine. With robust support for containerization and environment management, CWL ensures reproducibility and scalability in scientific workflows. By leveraging container technologies like Docker, it facilitates the encapsulation of software dependencies, making it effortless to share and reproduce workflows across diverse computing environments. Additionally, CWL incorporates a built-in secret module that provides secure management of sensitive information within workflows, allowing researchers to handle confidential data while adhering to privacy and security requirements. The CWL ecosystem expands beyond $\texttt{cwltool}$ and encompasses other software tools such as $\texttt{Arvados}$, $\texttt{Toil}$, and $\texttt{StreamFlow}$, which extend CWL's capabilities to remote platforms. These tools enable the execution of workflows on distributed computing resources and cloud infrastructures. Supported by an active and collaborative community, CWL continues to grow, offering a repository of pre-built workflows.\\

\begin{figure}[H]
    \centering
    \begin{minipage}{0.48\textwidth}
        \includegraphics[width=\textwidth]{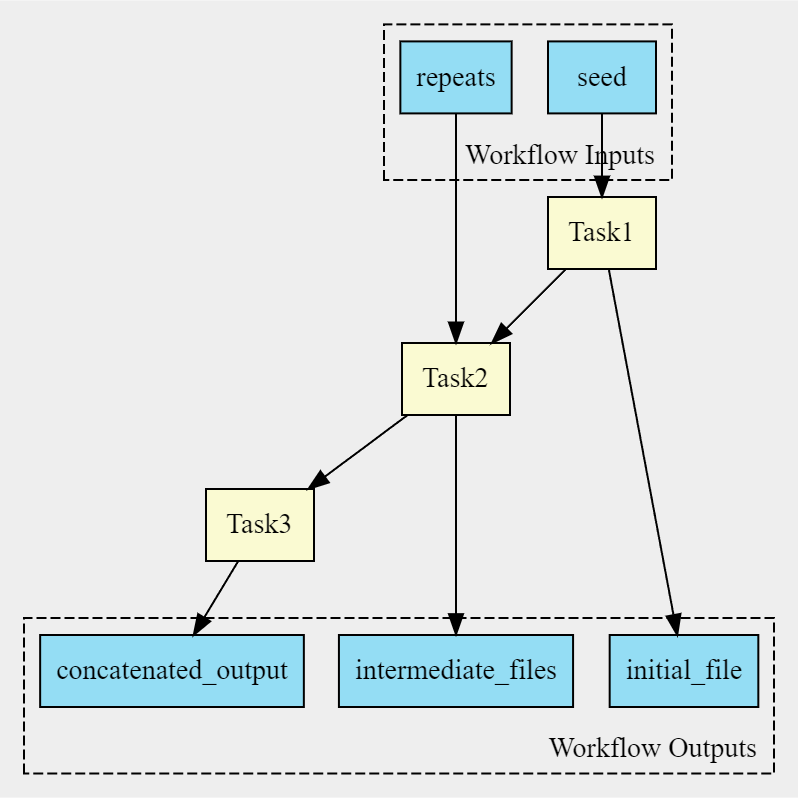}
    \end{minipage}\hfill
    \begin{minipage}{0.5\textwidth}
        \includegraphics[width=\textwidth]{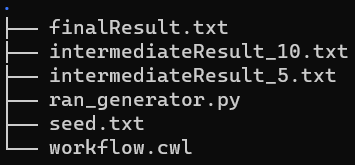}
    \end{minipage}
    \caption{Directed Acyclic Graph (DAG) generated by an external Graphviz program with $\texttt{--print-dot}$ flag and output directory structure for simple workflow in CWL.}
    \label{fig:demoCWL}
\end{figure}

Figure \ref{fig:demoCWL} shows the implementation of the example workflow. The whole workflow is stored in the file \textit{workflow.cwl}. CWL will automatically generate a temporary directory for all the tasks and only keep the required outputs unless specified with optional commands, therefore there is no need to specify the directories for different steps. 

\begin{python}
# workflow.cwl
cwlVersion: v1.2
class: Workflow
requirements:
  ScatterFeatureRequirement: {}
  InlineJavascriptRequirement: {}
inputs:
  seed:
    type: int
    default: 42
  repeats:
    type: int[]
    default: [5, 10]
outputs:
  initial_file:
    type: File
    outputSource: Task1/seed_file
  intermediate_files:
    type: File[]
    outputSource: Task2/intermediate_files
  concatenated_output:
    type: File
    outputSource: Task3/concatenated_output

steps:
  Task1:
    in:
      seed: seed
    out: [seed_file]
    run:
      class: CommandLineTool
      baseCommand: echo
      stdout: seed.txt
      inputs:
        seed:
          type: int
          inputBinding:
            position: 1
      outputs:
        seed_file:
          type: File
          outputBinding:
            glob: seed.txt

  Task2:
    in:
      seed_file: Task1/seed_file
      NumberOfRandoms: repeats
    scatter: NumberOfRandoms
    out: [intermediate_files]
    run:
      class: CommandLineTool
      requirements:
        DockerRequirement:
          dockerPull: 'localhost:5000/localRegistry'
        InitialWorkDirRequirement:
          listing:
          - entryname: ran_generator.py
            entry:
              $include: ran_generator.py
      baseCommand: ["python3", "ran_generator.py"]
      stdout: intermediateResult_$(inputs.NumberOfRandoms).txt
      inputs:
        seed_file:
          type: File
          inputBinding:
            position: 1
        NumberOfRandoms:
          type: int
          inputBinding:
            position: 2
      outputs:
        intermediate_files:
          type: File
          outputBinding:
            glob: intermediateResult_$(inputs.NumberOfRandoms).txt

  Task3:
    in:
      intermediate_files: Task2/intermediate_files
    out: [concatenated_output]
    run:
      class: CommandLineTool
      baseCommand: cat
      stdout: finalResult.txt
      inputs:
        intermediate_files:
          type: File[]
          inputBinding:
            position: 1
      outputs:
        concatenated_output:
          type: File
          outputBinding:
            glob: finalResult.txt
\end{python}

\begin{python}
#random generator for intermediate stage: ran_generator.py
#!/usr/bin/env python
import numpy
import sys

input_file = str(sys.argv[1])
number_of_randoms = int(sys.argv[2])

with open(input_file, "r") as input:
    numpy.random.seed(int(input.readlines()[0]))
for i in range(number_of_randoms):
    print(str(numpy.random.random()))
\end{python}

To execute Task2 using the pre-built container image ``container:v1'', it is necessary to manually push the docker image ``container:v1'' to the local registry named ``localRegistry''. We then call the workflow from the command line with $\texttt{cwltool workflow.cwl}$.\\

\noindent\textsc{User Interface}
\begin{itemize}
\item \textbf{Workflow language}:
YAML-based specification with additional support for inline JavaScript.
\item \textbf{Relation to analysis code}:
The description of the workflow factorizes entirely from the analysis code (separate scripts). But it is also possible to partially integrate the analysis codes into the workflow script.
\item \textbf{Boilerplate code}:
A lot. The specification of inputs and outputs has to be repeated multiple times: More than once (inputs, outputs and stdout) in each process, once in each step and once for the entire workflow. The input position of each variable in the command line requires manual configuration. Invoking external scripts and accessing external data are also quite complicated.
\item \textbf{Visualization and monitoring}:
Built-in tracing and visualisation tools are partially provided. However, an external tool for DAG visualisation is necessary. Dynamic workflow monitoring and dynamic DAG are not supported. Partial results and log files can be inspected manually for running and completed tasks.
\item \textbf{Learning curve}:
Hard to start or customise compared to Nextflow. However, the learning process is still quicker than with Yadage, thanks to its more complete tutorials, documentation and much larger user community.
\end{itemize}
\textsc{Features}
\begin{itemize}
\item \textbf{Supported programming languages}:
The system is agnostic to the programming language used for the implementation of steps.
\item \textbf{Data formats}:
Any formats. No specific features for the inspection of data formats.
\item \textbf{Dependency management}: Centralized in \textit{workflow.cwl} where the dependencies are manually specified in each step. One-to-many, many-to-one, and many-to-many dependencies are possible by defining array inputs and outputs. Dynamical, conditional and looping dependencies can be implemented with built-in modules.
\item \textbf{Execution control}: CWL does not support pausing or resuming workflows,
and it does not allow changes to be made during execution either. A partial workflow can be executed using the target option. When specifying a target as an output parameter, only the steps contributing to that output will be executed. If a target is set as a workflow step, the execution will start from that specific step. Similarly, if a target is defined as an input parameter, only the steps connected to that input will be run.
\item \textbf{Error handling}:
The workflow will be marked as failed if any of its individual tasks fails. Error messages produced during the execution of failed tasks are reported. However, (incomplete) outputs of failed tasks are not deleted automatically.
\item \textbf{Logging and provenance}:
The specification of log files is not allowed. Error messages with traceback information are recorded in the execution history logs and can be manually exported.
\item \textbf{Version control}:
The code used to create output files can be tracked in log files. External version control tools like git can also be used for additional tracking and management of code changes.
\item \textbf{Scalability}:
Complicated scalability. It supports local parallel executions on multiple CPU cores. However, for remote executions on different platforms, the use of various software that supports CWL is required.
\end{itemize}
\textsc{Resource Integrations}
\begin{itemize}
\item \textbf{Software and environment management}:
CWL has a built-in software management tool to specify software requirements. Docker container and singularity can be executed as a task.
\item \textbf{Storage systems support}:
Support both local and remote storage systems. 
\item \textbf{Remote execution system support}: 
The built-in $\texttt{cwltool}$ only supports local execution. However, other software such as $\texttt{Arvados}$, $\texttt{Toil}$ and $\texttt{StreamFlow}$ provide support for remote platforms such as AWS, Azure, GCP, Grid Engine, HTCondor, LSF, Mesos, OpenStack, Slurm, PBS/Torque, Kubernetes, HPC with Singularity, Occam and more. 
\item \textbf{Authentication and authorization mechanisms}:
CWL provides a built-in secret module that stores sensitive information such as passwords, access tokens, and encryption keys securely. 
\end{itemize}
\textsc{Installation and Configuration}
\begin{itemize}
\item \textbf{Installation}:
Python3 is necessary for using CWL. CWL can be easily installed with Conda or pip without root access rights. 
\item \textbf{Architecture}:
Multiple applications on a single machine. Upon restart, existing output files will not be detected and all tasks will be re-run. 
\item \textbf{State management}:
Target based.
\item \textbf{Portability}:
Workflows can be easily executed from different locations with the help of containers and singularity.
\end{itemize}
\textsc{Support and Management of the Tool}
\begin{itemize}
\item \textbf{Documentation}: 
Sparse and incomplete documentation on the official website. Open source code on Github.
\item \textbf{Support}:
Extensive support on stackoverflow.com, Discourse Group and Matrix Space. Weekly and monthly community meetings are also officially organised. CWL is designed to meet the needs of data-intensive science, such as bio-informatics, medical imaging, astronomy, high energy physics, and machine learning.
\item \textbf{Tool developers}:
A community with oversight from the CWL leadership team.
\item \textbf{History and project activity}:
CWL exists since 2016 with active ongoing development (about 10 commits monthly).
\item \textbf{User community}:
Active and huge community. Designed for different fields of data-intensive science but mainly used by bio-informatics communities.
\item \textbf{Long term perspective}:
Good perspective, given active development and community.
\item \textbf{Lock-in}:
No. There is no need to modify the individual analysis code before implementing CWL. Some other workflow management systems, such as Snakemake and REANA, also provide support for CWL.
\item \textbf{License}:
Nextflow is licensed under the Apache 2.0 License with free permission to alter it in any way.
\item \textbf{Use in PUNCH}: Astrophysics, ALICE, ATLAS, CMS and LHCb.
\end{itemize}

\subsection{Workflow Management Platforms}
Workflow Management Platforms help to containerize computational environments and to run workflows on remote computing clouds. The selection of reviewed workflow management platforms is influenced by the authors' background and can be extended in future.
\subsubsection*{Snakemake workflows on the Reproducible research data analysis platform (Reana)\footnote{https://github.com/reanahub/reana}}
Reana serves as a platform that provides workflow execution as a service and is agnostic to the utilized workflow framework. We here discuss the execution of snakemake workflows on Reana servers, for which not all previously mentioned features of snakemake are available.\\

\begin{figure}[h]
    \begin{minipage}{0.1\textwidth}
    \centering\includegraphics[width=\textwidth]{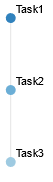}
    \end{minipage}\hfill
    \begin{minipage}{0.45\textwidth}
        \includegraphics[width=\textwidth]{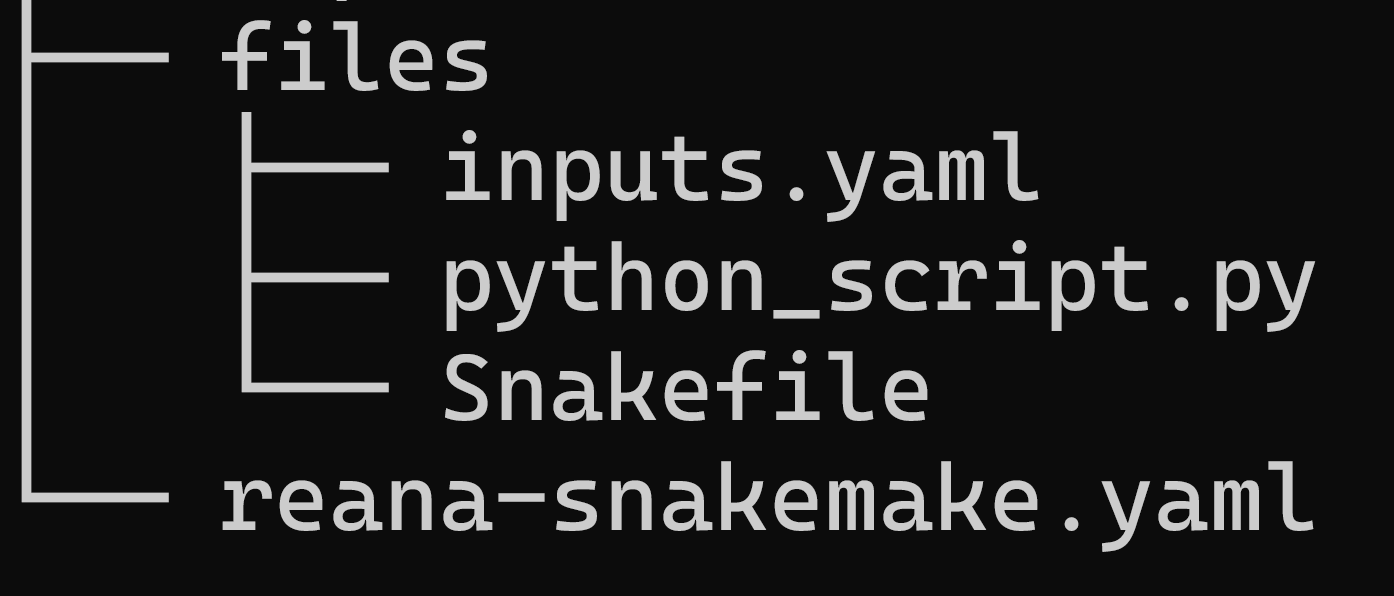}
    \end{minipage}
    \caption{Directed Acyclic Graph (DAG) taken from Reana report and directory structure for simple workflow in Reana.}
    \label{fig:demoReana}
\end{figure}

Figure \ref{fig:demoReana} illustrates the implementation of the example workflow. Task 1 generates a random seed. It is used by Task 2 to generate random numbers, by calling a separate python script. Since Reana does not allow to directly pass arguments to scripts (only \texttt{shell} directives are supported from snakemake), additional parsing code is necessary. With each call of Task 2, we pass a wildcard parameter to the script, which specifies how many numbers are generated. Task 3 concatenates the outputs of the calls of Task 2. Reana does not support the conda package management in Snakemake, but containers can be spawned from given images.

\begin{python}
#Snakefile
rule Task3:
    input:
        "intermediateResult_10.txt",
        "intermediateResult_5.txt"
    output:
        "finalResult.txt"
    shell:
        "cat {input} > {output}"

rule Task2:
    input:
        seed = "seed.txt",
        script = config["generate_randoms"]
    output:
        "intermediateResult_{NumberOfRandoms}.txt"
    shell:
        "python {input.script} {input.seed} {wildcards.NumberOfRandoms} {output}"

rule Task1:
    output:
        "seed.txt"
    shell:
        "echo 42 > {output}"
\end{python}
\begin{python}
#inputs.yaml
generate_randoms: files/python_script.py
\end{python}
\begin{python}
#reana-snakemake.yaml
inputs:
   directories:
     - files
   parameters:
     input: files/inputs.yaml
workflow:
   type: snakemake
   file: files/Snakefile
outputs:
   files:
     - finalResult.txt
\end{python}
\begin{python}
#python_script.py
import numpy

import sys
input_file = str(sys.argv[1])
number_of_randoms = int(sys.argv[2])
output_file = str(sys.argv[3])

with open(input_file,"r") as input:
    numpy.random.seed(int(input.readlines()[0]))
with open(output_file,"w") as output:
    for i in range(number_of_randoms): output.write(f"{numpy.random.random()}\n")
\end{python}
\textsc{User Interface}
\begin{itemize}
\item \textbf{Workflow language}:
Simplistic built-in workflow management framework called Serial, which permits to run tasks only in a linear sequential manner (therefore not reviewed here). Support for Common Workflow Language, Yadage and Snakemake workflows.
\item \textbf{Relation to analysis code:}
Depending on chosen workflow language. For snakemake workflows, full factorization of workflow logic and analysis code.
\item \textbf{Boilerplate code}:
Yaml file to specify computing environment, e.g. workflow, workflow language, input and output files etc. For snakemake workflows, parameters cannot directly be passed to scripts and additional .yaml files and parsing code are required.
\item \textbf{Visualization and monitoring}:
Dynamic monitoring and static DAG visualization.
\item \textbf{Learning curve}:
Simple.
\end{itemize}
\textsc{Features}
\begin{itemize}
\item \textbf{Supported programming languages}:
Depending on chosen workflow language. Multiple for snakemake workflows, but scripts cannot be called directly and are instead invoked via shell commands.
\item \textbf{Data formats}:
Depending on chosen workflow language; generally any format.
\item \textbf{Dependency management}:
Depending on chosen workflow language. For snakemake, one-to-many, many-to-one, and many-to-many dependencies.
\item \textbf{Execution control}:
Workflows cannot be paused. Changes require a restart.
\item \textbf{Error handling}:
Each execution requires a new upload and runs in a clean environment.
\item \textbf{Logging and provenance}:
Individual log files for all tasks are accessible at run-time. For snakemake workflows, additional global log files for each workflow execution, including the executed tasks with their parameters, timestamps and activated environments. Benchmarking in reana report.
\item \textbf{Version control and archivability}:
Previously uploaded workflows are accessible on server.
\item \textbf{Scalability}: Simple. Apart from the built-in Serial framework, all supported workflow management frameworks support parallel execution.
\end{itemize}
\textsc{Resource Integration}
\begin{itemize}
\item \textbf{Software and environment management}:
Workflows are executed in containerized environments on server.
\item \textbf{Remote execution system support}: 
Support for HTCondor, Kubernetes and Slurm compute backends.
\item \textbf{Storage systems support}:
Support for GitLab, CVMFS, EOS and NFS storage systems.
\item \textbf{Authentication and authorization mechanism}:
Users are authenticated using private access tokens to upload workflows to server.
\end{itemize}
\textsc{Installation and Configuration}
\begin{itemize}
\item \textbf{Installation}:
Python and a web browser are necessary for using Reana. Reana-client can be easily installed from the Python Package Index (pip), without root access rights. Server can be selected from Reana cluster or installed individually.
\item \textbf{Architecture}:
Client-server architecture. Once the workflow is uploaded, the execution cannot be terminated.
\item \textbf{State management:} 
Target based.
\item \textbf{Portability}:
Workflows are executed on server.
\end{itemize}
\textsc{Support and Management of the Tool}
\begin{itemize}
\item \textbf{Documentation}: Sparse and incomplete documentation on the official website. Open source code on Github.
\item \textbf{Support}:
Reana forum, chat on Mattermost.
\item \textbf{System developers}:
Developed by CERN.
\item \textbf{History and project activity}:
Established in 2017. Active with multiple new commits per month.
\item \textbf{User community}:
Originally developed for high energy physics, but translates to any research field.
\item \textbf{Long term perspective}:
Good, given active development and support by CERN.
\item \textbf{Lock-in}:
Unlikely, due to support for different workflow management frameworks. However, e.g. snakemake workflows might need to be adjusted to bypass snakemake features that are not supported on Reana.
\item \textbf{License}:
CERN Copyright.
\item \textbf{Use in PUNCH}: ALICE, ATLAS, CMS and LHCb
\end{itemize}
\begin{sidewaystable}[h]
\begin{center}
\begin{minipage}{\textwidth}
\caption{Evaluation of workflow management system assessment criteria on selected technical solutions with \textcolor{green}{advantages} and \textcolor{red}{disadvantages}. Part I}
\label{tab:assessment1}
\begin{tabular}{l|llll}
& Criteria & Luigi & Snakemake & Yadage\\
\hline

\parbox[t]{2mm}{\multirow{5}{*}{\rotatebox[origin=c]{90}{Interface}}} 

& Workflow language & Python  & custom Python-based & YAML-based\\
& Relation to analysis code &  \textcolor{red}{integrated}  & \textcolor{green}{complete factorization} & \textcolor{green}{complete factorization}  \\
& Boilerplate code &  minimal &  minimal & \textcolor{red}{a lot}\\
& Visualization and monitoring & \textcolor{green}{extensive}  &  \textcolor{red}{no dynamic DAG} &  \textcolor{red}{no dynamic DAG} \\
& Learning curve & \textcolor{red}{intermediate} & \textcolor{green}{simple}  & \textcolor{red}{hard}\\\hline

\parbox[t]{2mm}{\multirow{7}{*}{\rotatebox[origin=c]{90}{Features}}} 

& Programming languages & \textcolor{red}{single}  & \textcolor{green}{multiple}  &  \textcolor{green}{multiple} \\
& Data formats & any  &  any  &  any \\
& Dependency management & \textcolor{green}{extensive}  &  \textcolor{green}{extensive}  & \textcolor{red}{incomplete}\\
& Execution control &  extensive &  extensive  &   \textcolor{red}{very limited}\\
& Error handling, provenance, logging &  \textcolor{red}{failed output remains}  &  \textcolor{green}{failed output deleted}  & \textcolor{red}{failed output remains} \\
& Version control and archivability & external & external and internal & external \\
& Scalability & easy  &  easy  &  easy \\\hline

\parbox[t]{2mm}{\multirow{4}{*}{\rotatebox[origin=c]{90}{Resources}}}

& Environment management & some support & \textcolor{green}{extensive support}  & some support  \\
& Storage systems support &  extensive &  extensive &  \textcolor{red}{limited} \\
& Remote execution support & extensive  &  extensive  &  \textcolor{red}{limited} \\
& Authentication mechanism & environment variables  & environment variables  & environment variables\\\hline

\parbox[t]{2mm}{\multirow{4}{*}{\rotatebox[origin=c]{90}{Install}}}

& Installation & pip, non-root  &  conda, non-root  & pip, non-root  \\
& Architecture & single application  &  single application  &  single application \\
& State management &  target based & target based  &   target based \\
& Portability &  easy &  easy  &  easy \\\hline

\parbox[t]{2mm}{\multirow{9}{*}{\rotatebox[origin=c]{90}{Support}}}

& Documentation & extensive  & extensive  &  \textcolor{red}{fragmentary} \\
& Support & extensive  & extensive  &  \textcolor{red}{minimal} \\
& System developers & Spotify Group  &  academic team  &  \textcolor{red}{tiny academic team} \\
& History and activity & very active  &  very active &  \textcolor{red}{very inactive} \\
& User community & large  & large  &  \textcolor{red}{minimal} \\
& Long term perspective &  good & good  &  \textcolor{red}{bad} \\
& Lock-in & \textcolor{red}{yes}  & \textcolor{green}{no}  &  \textcolor{green}{no} \\
& License & Apache 2.0, free use  &  MIT, free use&  MIT, free use \\
& Use in PUNCH & CMS, Belle 2 & LHCb, Radioastronomy & ALICE, ATLAS, CMS, LHCb\\\hline

\end{tabular}
\end{minipage}
\end{center}
\end{sidewaystable}

\begin{sidewaystable}[h]
\begin{center}
\begin{minipage}{\textwidth}
\caption{Evaluation of workflow management system assessment criteria on selected technical solutions with \textcolor{green}{advantages} and \textcolor{red}{disadvantages}. Part II}
\label{tab:assessment2}
\begin{tabular}{l|l|ll||l}
& Criteria & Nextflow & CWL & Reana \\
\hline

\parbox[t]{2mm}{\multirow{5}{*}{\rotatebox[origin=c]{90}{Interface}}} 

& Workflow language & Groovy-based  &  YAML-based & Python-based \\
& Relation to analysis code &  \textcolor{green}{complete factorization}  & \textcolor{green}{complete factorization}  & \textcolor{green}{complete factorization}  \\
& Boilerplate code &  minimal  &   \textcolor{red}{a lot} & some\\
& Visualization and monitoring  &  \textcolor{red}{no dynamic DAG} &  \textcolor{red}{incomplete} & \textcolor{red}{no dynamic DAG} \\
& Learning curve  & moderate  &  \textcolor{red}{hard} & \textcolor{green}{simple}\\\hline

\parbox[t]{2mm}{\multirow{7}{*}{\rotatebox[origin=c]{90}{Features}}} 

& Programming languages & \textcolor{green}{multiple}  & \textcolor{green}{multiple}  &  \textcolor{green}{multiple} \\
& Data formats & any  &  any  &  any \\
& Dependency management &  \textcolor{green}{extensive}  &  \textcolor{green}{extensive}  &  \textcolor{green}{extensive} \\
& Execution control &  extensive  &  \textcolor{red}{limited} &  \textcolor{red}{limited}\\
& Error handling, provenance, logging & \textcolor{red}{failed output remains}  & \textcolor{red}{failed output remains} & \textcolor{green}{single-use clean environment}\\
& Version control and archivability & external and internal & external and internal & external and internal \\
& Scalability &  easy  &  \textcolor{red}{complicated} & easy \\\hline

\parbox[t]{2mm}{\multirow{4}{*}{\rotatebox[origin=c]{90}{Resources}}}

& Environment management & \textcolor{green}{extensive support}  & some support & some support  \\
& Storage systems support &  extensive &  extensive &  extensive \\
& Remote execution support &  \textcolor{green}{maximal}  &  \textcolor{red}{complicated} &  extensive \\
& Authentication mechanism & \textcolor{green}{centralized and secured}  &  \textcolor{green}{secured} &  access tokens\\\hline

\parbox[t]{2mm}{\multirow{4}{*}{\rotatebox[origin=c]{90}{Install}}}

& Installation &  wget, curl, conda, non-root  & pip, conda, non-root & pip, non-root \\
& Architecture &  single application  &  \textcolor{red}{multiple applications} & single application  \\
& State management &  target based & target based  &   target based \\
& Portability &  easy &  easy  &  easy \\\hline

\parbox[t]{2mm}{\multirow{9}{*}{\rotatebox[origin=c]{90}{Support}}}

& Documentation & extensive  &  \textcolor{red}{incomplete} & \textcolor{red}{incomplete}\\
& Support & \textcolor{green}{excellent}  &  satisfactory &  satisfactory\\
& System developers &  Seqera Labs  &  large community &  CERN \\
& History and activity &  very active &  active & \textcolor{red}{less active}\\
& User community & significant  &  large &  significant \\
& Long term perspective & good  &  good &  acceptable\\
& Lock-in & \textcolor{green}{no}  & \textcolor{green}{no}  &  \textcolor{green}{no} \\
& License &  Apache 2.0, free use&  Apache 2.0, free use &  CERN \\
& Use in PUNCH & \textcolor{red}{None} & Astrophysics, ALICE  & ALICE, ATLAS, CMS, LHCb \\
&&&ATLAS, CMS, LHCb&\\
\hline

\end{tabular}
\end{minipage}
\end{center}
\end{sidewaystable}

\section{Recommendations}
\label{recommendations}
Before looking at specific workflow management systems we would like to give the following general recommendations:
\begin{enumerate}
\item If your workflow does or will in future consist of more than three interdependent steps, use a workflow management system.
\item Check your requirements. Which workflow management system features are essential, nice to have, irrelevant for the problem you want to solve.
\item Is there an established workflow management system in your community? If yes, check if it fulfill your requirements.
\end{enumerate}

Any recommendations of specific tools will naturally depend on the selection of the reviewed workflow management systems, of which a large variety exists.
The current selection is certainly influenced by the background of the authors.
We explicitly welcome extensions of the comparison to more workflow systems.

When choosing a workflow management system, the relevance of the above criteria has to be assessed case by case.
Also, a possible lock-in should be considered because requirements can change and may call for a change of the workflow management system.
Of course, the prevalent use of a particular system in a specific research field may play a significant role.

Based on the criteria above, we recommend the use of Snakemake as workflow management system in research of PUNCH unless the general recommendations suggest something else.
Snakemake comes with extensive support for environment management and remote execution.
At the same time, it features a simple but powerful syntax that allows for a complete factorization of workflow logic and analysis code.
Therefore, the implementation of existing analyses in a workflow is possible with minimal effort and lock-in is very unlikely.
It supports multiple programming languages and Snakemake workflows can be run on the Reproducible Research Data Analysis Platform (Reana) for extended functionalities.

Another notable contender is Nextflow. In comparison to Snakemake, Nextflow requires additional code for organizing steps within a workflow and manually specifying their dependencies. Java is necessary instead of Python, which could pose a challenge for certain computing systems. However, it offers powerful features that simplify the code. It provides flexibility with support for pausing and resuming executions, allowing changes to be made before resuming without the need for a complete restart. Nextflow also prioritizes the secure handling of sensitive information by centralizing and safeguarding authentication and authorization mechanisms with its built-in module. Furthermore, Nextflow benefits from collaborations with major tech companies like Meta, Google, Amazon, and Microsoft, providing easy access to a wide range of computing services. While the user community is large and active, currently it does not include PUNCH.

In this document we focus on workflows that do not have special requirements and can run on linux systems with decent software support, e.g. availability of Python, Java, container management. Some workflows might be tightly integrated with specific hardware or software environments and tools optimized for such environments are often more suited than generic workflow management systems. An example in PUNCH are lattice QCD calculations running on supercomputers. Here SLURM is often a better choice because specific features such as the management of task arrays or multi-clusters are usually not supported by general-purpose tools which would lead to a suboptimal utilization of computing resources and problems with distributed filesystems and long workflow execution times are avoided.

\section{Conclusions}
\label{conclusions}
The use of workflow management systems should become a standard in physics research in the interest of transparency, reproducibility and data preservation.
This has also immediate advantages for the individual analyzer, because it reduces the risk of (undetected) mistakes and saves time and tedious labor if analyses or parts of it have to be rerun several times.

Many workflow systems with different features are available and the choice of a system can become overwhelming.
We compared selected workflow management systems in regards of a number of criteria relevant in physics research, summarized in Tables~\ref{tab:assessment1} and \ref{tab:assessment2}, to help with the choice of workflow management.

\backmatter

\bmhead{Acknowledgments}
We thank Lukas Heinrich for helpful comments. We also thank Daniel Knüttel for sharing his experience in utilizing Nextflow and SLURM on supercomputers for lattice QCD calculations. We acknowledge support by the Particle, Universe, Nuclei and Hadrons consortium of the NFDI (PUNCH4NFDI).

\end{document}